\documentclass[12pt]{article}
\pdfoutput=1
\usepackage[utf8x]{inputenc}
\usepackage{amssymb}
\usepackage{amsmath}
\usepackage{epsfig}
\usepackage{graphicx}
\usepackage{color}
\usepackage{ulem}

\makeatletter

\@addtoreset{equation}{section}
\makeatother
\newcommand{\be}{\begin{equation}}
\newcommand{\ee}{\end{equation}}
\newcommand{\ben}{\begin{equation*}}
\newcommand{\een}{\end{equation*}}
\newcommand{\bea}{\begin{eqnarray}}
\newcommand{\eea}{\end{eqnarray}}
\newcommand{\bag}{\begin{align}}
\newcommand{\eag}{\end{align}}

\newcommand{\bx}{{\bf x}}

\newcommand{\al}{\alpha}

\begin{document}

\begin{titlepage}
\vskip1cm
\begin{flushright}
\end{flushright}
\vskip1.25cm
\centerline{
\bf \large Patterns of Gravitational Cooling in Schr\"{o}dinger Newton  System} 
\vskip1cm \centerline{ \textsc{
Dongsu Bak,$^{\tt a,c}$ Seulgi Kim,$^{\tt a,c}$   Hyunsoo Min,$^{\tt a}$ Jeong-Pil Song$^{\, \tt b}$} }
\vspace{1cm} 
\centerline{\sl  a) Physics Department,
University of Seoul, 
Seoul 02504 \rm KOREA}
 \vskip0.3cm
 \centerline{\sl b) Department of Chemistry, Brown University, Providence, 
 RI 02912 \rm USA}
 \vskip0.3cm
\centerline{\sl  c) Natural Science Research Institute,
University of Seoul, 
Seoul 02504 \rm KOREA}

 \centerline{
\tt{(dsbak,\,hsmin@uos.ac.kr,\,seulgi.kim@cern.ch,\,jeong\_pil\_song@brown.edu})}
  \vspace{1cm}

\centerline{ABSTRACT} \vspace{0.75cm} \noindent
{ 
We study time evolution of  Schr\"{o}dinger-Newton system 
using the self-consistent Crank-Nicolson method to 
understand the dynamical characteristics of nonlinear systems.
Compactifying the radial coordinate 
by  a new one, which brings the spatial infinity to a finite value,  we are able to impose the boundary 
condition at infinity 
 allowing for a numerically exact treatment of the Schr\"{o}dinger-Newton equation.
We study patterns of gravitational cooling  starting from exponentially  localized initial states. When the gravitational attraction is strong enough,
we find that a small-sized oscillatory solitonic core   is forming quickly, which is  surrounded by
a growing 
number of temporary halo states.  In addition
a significant fraction of particles escape  to 
asymptotic regions.   The system eventually settles down to a stable solitonic core state while all the excess kinetic energy is carried 
away by the escaping 
particles, which is  a phenomenon of gravitational cooling. 
}
\end{titlepage}



\section{Introduction}
The  Schr\"{o}dinger-Newton  system \cite{Diosi:2014ura, Moroz:1998dh,Bahrami:2014gwa}
is an interesting example of gravitational nonlinear system, which has been studied
through numerical analysis.   It has 
 a wide range of applications to physics
including studies of the measurement problem \cite{vanMeter:2011xr} and the
 dark matter problem of our Universe \cite{Hui:2016ltb,Lee:2017qve,Schive:2014dra}. 
 
 Its numerical studies 
 and the effect of artificial truncation for the 
treatment of computational domains 
turn out to be subtle due to the boundary condition imposed at infinity
 as  a significant fraction of particles escape  to asymptotic regions for a set of rather generic initial conditions.  Thus to be accurate,
one has  to impose the boundary condition exactly in a numerically reliable manner
to avoid the boundary problem arising from the removal of the computational domains.
To deal with this issue, a choice adopted in the literature \cite{sponge}
 is to use the so-called sponge boundary condition at some large radial position where one imposes a purely absorbing boundary condition. However 
 this involves approximations and it is not clear how the stability of numerical analysis can be maintained with  the nonlinear nature of the system.  
Another alternative is to use a periodic boundary condition but in this case the matter escaping to asymptotic regions are returning to center 
due to the artificial periodic boundary condition leading to  a rather serious  finite size effect in some cases. 

In this note, we shall follow a different method \cite{Bak:2011ga}
  based on a compact coordinate $Y$ ranged over $[0,1)$ defined  by the relation $Y=\frac{r}{1+r}$  where $r$ is 
the radial coordinate 
 ranged over $[0, \infty)$.  So  $r=\infty$ is mapped to $Y=1$ for instance. Thus at least in principle the boundary condition can be imposed at $Y=1$ in a 
 straightforward manner. The other part is to use the self-consistent Crank-Nicolson algorithm to see the time evolution of the system starting with a set of initial wave functions which  are localized around $Y=0$. We test our method in a various manner. When the gravity is turned off, we shall compare our result against the corresponding 
 exact solution. In addition, 
we shall find the relevance of our method with an 
emphasis on conserved quantities 
as we shall demonstrate below. 
 
 We then study the pattern of gravitational cooling of the system starting with  initial condition in which the wavefunction is exponentially decaying in $r$ coordinate. 
 When the total mass is larger than a certain critical value, one finds that a small-sized solitonic core 
 forms as a result of gravitational attraction together with 
 a growing 
number of excited halo states while a significant fraction of particles escapes to asymptotic 
 regions. Eventually all the excited halo states decay away leaving behind only the stable solitonic core state, which is nothing but 
the phenomenon of gravitational 
 cooling \cite{Guzman:2004wj,Guzman:2006yc}. In this process, any excess kinetic energy will be released to asymptotic regions with a free streaming flux of particles. In this note, we shall 
present detailed patterns appearing in this gravitational  relaxation process whose character does not seem to be found elsewhere.

\section{Schr\"{o}dinger Newton Gravity}\label{Sec2}
We shall begin with the Schr\"{o}dinger-Newton system \cite{Diosi:2014ura, Moroz:1998dh} described by
\begin{align}    \label{}
i\hbar \partial_t \psi(\bx,t ) & = -\frac{\hbar^2}{2m}\nabla^2 \psi(\bx,t )  +m V(\bx,t ) \psi(\bx,t )  
\\
 \nabla^2 V(\bx,t )&=4\pi GM \, |\psi|^2(\bx,t ) 
\end{align}
where the Newtonian potential $V$ satisfies the boundary condition $V \rightarrow 0$ as $r \rightarrow \infty$ and the wavefunction $\psi$  is normalized 
such that
\be
\int d^3 \bx\,  |\psi|^2 
=1 .
\label{proba}
\ee
This system consists of many Schr\"{o}dinger  particles of mass $m$ whose total mass is given by $M$. The wavefunction $\psi$ is describing a Bose-Einstein condensate of the constituent particles. These  Schr\"{o}dinger particles are self-interacting gravitationally, which  lead
to an integro-nonlinear equation when $V$ is represented by an integration in terms of $|\psi|^2$. 

In this system, the total number of particles $N=M/m$ is preserved in time. In addition, the total energy including the gravitation contribution is conserved in time.
One may easily show that this  conserved energy is given by
\bea
{\cal E}= - \frac{\hbar^2}{2m} \langle\nabla^2 \rangle  +\frac{m}{2}\langle  V \rangle .
\eea
On the other hand, the expectation of the Hamiltonian that is referred as Hamiltonian energy 
\bea
{E(t)}= \langle  H \rangle =- \frac{\hbar^2}{2m} \langle\nabla^2 \rangle  +m\langle  V \rangle 
\eea
 is time-dependent in general. Below we shall use the conserved total energy and the total probability (\ref{proba}) to show 
 the stability and effectiveness of our numerical analysis.

We shall work with dimensionless variables by the following 
rescaling 
\begin{align}    \label{}
t & \rightarrow   \al_t \, t   = \frac{\hbar^3}{m^3} \left(\frac{F}{4\pi GM}\right)^2 t \\
\bx &  \rightarrow   \al_s \, \bx   = \frac{\hbar^2}{m^2} \frac{F}{4\pi GM} \, \bx\\
\psi & \rightarrow   \al_\psi \, \psi   = \frac{m^3}{\hbar^3} \left(\frac{4\pi GM}{F}\right)^{\frac{3}{2}}  \psi \\
V& \rightarrow   \al_V \, V   = \frac{m^2}{\hbar^2} \left(\frac{4\pi GM}{F}\right)^{2}  V
\end{align}
leading to a form 
suited for the numerical analysis
\begin{align}    \label{SchEq}
i \partial_t \psi(\bx,t ) & = -\frac{1}{2}\nabla^2 \psi(\bx,t )  + V(\bx,t ) \psi(\bx,t )  
\\
 \nabla^2 V(\bx,t )&=\ F \, |\psi|^2(\bx,t ) \label{LapEq}
\end{align}
which is in terms of the above dimensionless variables.
Note that $\al_\psi$ and $\alpha_V$ can be given by relations $\al^2_\psi  \al_s^3 =1$ and $\al_V\al_t= \frac{\hbar}{m }$ respectively. 
Below we shall choose $m=10^{-22}\, \text{eV}/{c^2}$ and $F$ such that
\be
{F} =
 \frac{M}{10^7 M_{\odot}}
\ee
with  $M_{\odot}$ denoting the solar mass.  This then fixes the time and length scales as
  \begin{align}    \label{}
 \al_t & \sim 2.36 \times 10^7 \, \text{yr} \\
 \al_s & \sim 0.680\, \, \, \text{k\,pc}
\end{align}
when $M=10^7 M_{\odot}$. 
With $F/M$ fixed, the scales of our analysis will be fixed. Thus the total mass $M$ of the system will be traded by the parameter $F$ in our numerical 
analysis. 

In this note,  we shall mainly be
interested in a time evolution of this nonlinear  Schr\"{o}dinger  system. One can of course study an eigenvalue  problem of the stationary
version 
of the above  Schr\"{o}dinger  equation.  In the following, we shall  restrict our study for the spherically symmetric configuration for the simplicity of our numerical analysis. 
One is, however,  ultimately interested in problems with angular dependence and more complicated problems such as collision of lumps whose nature  is  truly $3+1$ dimensional. 
With the spherical
symmetry,
the eigenvalue problems can be solved leading to towers of stationary bound states even though the problems involve the nonlinearity.
 It turns out that all the excited states are unstable under a small perturbation of states 
and decay to a ground state\footnote{This ground state depends on $F$ together with the final remaining mass for the ground state which is obtained from $M$  by  
subtracting the mass that is ejected to asymptotic infinity at $t=\infty$.} together with an unbounded stream  
to asymptotic regions. This surely reflects the nonlinear nature of the problem since in a conventional linear  Schr\"{o}dinger  system,
this never happens unless there are some other perturbation in its Hamiltonian which derives  transitions from one state to another.
In our system, the transition occurs without any modification of the Hamiltonian and some significant 
portions
of particles are  always  ejected to asymptotic 
regions in this process of relaxation to the ground state. 

To see their general features, we shall consider two choices of the initial state
\bea
\psi_{I}(r, 0) &=& \left(\frac{a_{I}}{\pi}\right)^{\frac{3}{4}} e^{-\frac{a_{I}}{2} r^2} \cr
\psi_{II}(r, 0) &=& \left(\frac{a_{II}^3}{8\pi}\right)^{\frac{1}{2}} e^{-\frac{a_{II}}{2}r} 
\label{initial}
\eea
which are labeled by $I$ and $II$ respectively. 
Since the choice of $a_{I,II}$ can be traded with an additional rescaling of $F$, we shall fix $a_{I,II}=1$ for the definiteness, which is  without loss of generality. 

With these initial conditions, we shall numerically study the time evolution 
of the above equations. Roughly speaking, some part of the 
initial state will settle down to  a ground state while all the remaining will be eventually ejected to asymptotic regions. The final settlement of the ground state
is stable under a small state perturbation, which behaves as a kind of soliton\footnote{In some literatures, this state is called as a soliton star.}. 
As we shall see, in this process, the innermost core state is rapidly oscillating in general which is 
surrounded
by halos of excited states\footnote{These excited states differ from the ones in the above obtained by solving eigenvalue problems since here these states are affected 
by the presence of the core.} and the oscillation of the core  together with halos will eventually decay away either to the ground state or to the asymptotic regions. In this note, 
we would like to see the details 
of the time evolution as we change the parameter $F$.

\section{Numerical Setup}
Assuming a spherical symmetry, \eqref{SchEq} and \eqref{LapEq} become
\begin{align}
i \partial_t \psi(r,t ) & = -\frac{1}{2}\frac{1}{r^2}\partial_r\left(r^2 \partial_r\psi(r,t)\right)+V(r,t)\psi(r,t) \label{radSchEq} 
\\
 \frac{1}{r^2}\partial_r\left(r^2 \partial_r V(r,t)\right)&=\ F \, |\psi|^2(r,t) . \label{radLapEq}
\end{align}
We impose a Neumann boundary condition $\partial_r\psi(0,t)=0$ at $r=0$ and a Dirichlet boundary condition $\psi(r,t)=0$ at $r=\infty$ to have a finite value of the integral $4\pi \int_0^\infty |\psi|^2 r^2 dr $. However this boundary condition at $r=\infty$ is difficult to impose in a numerical analysis of the system. In many cases, people usually impose the Dirichlet condition at $r=R$ with sufficiently large value of $R$, which introduces unwanted reflection waves at the boundary. To resolve this problem, for instance in \cite{sponge}, a sponge function
has been used in finding stationary solutions of \eqref{radSchEq} and \eqref{radLapEq} 
for removing computational domains. 
In this work, we employ a different way, since we are interested in outgoing waves in time evolution. We compactify the region $0<r<\infty$ into a unit interval by introducing a new coordinate variable $Y=\frac{r}{1+r}$. Then there is no issue of reflection waves.
In terms of the new coordinate $Y$, the radial Laplacian is expressed as
\begin{align}
\nabla^2=\frac{1}{r^2}\partial_r\left(r^2 \partial_r\right)=\frac{(1-Y)^4}{Y^2}\partial_Y\left( Y^2\partial_Y\right) . \label{radLaplacian}
\end{align}

We discretize the spatial interval $0\leq Y \leq 1$ into $N_Y$ small intervals of the size $\Delta Y=1/N_Y$. Then we approximate $Y$ by $j \Delta Y $, $j=0,1,\cdots, N_Y$. The temporal interval is $0\leq t \leq t_f$. This time interval is discretized $N_t$ times with small intervals of the size  $\Delta t =t_f/N_t$. In this way we can approximate the time $t$ by
$t=k \Delta t $, $k=0,1,2,\cdots, N_t$. The scalar field $\psi(Y,t)$ and the potential function $V(Y,t)$ are replaced with the site variable $\psi_j^k$ and $V_j^k$. The first derivative and the second derivatives of
them are replaced by
\begin{align}
\partial_Y \psi \to \frac{\psi_{j+1}^k- \psi_{j-1}^k}{2\Delta Y}, \qquad
\partial^2_Y \psi \to \frac{\psi_{j+1}^k+ \psi_{j-1}^k-2\psi_{j}^k}{\Delta Y^2}
\end{align}
for $j= 1, 2, \cdots, N_Y-1$. On the other hand for $j=0$, this replacement becomes singular. Instead, we use the relation 
\begin{align}
\nabla^2 \psi|_{r=0}= 3 \partial^2_r \psi |_{r=0} =3[(1-Y)^4 \partial^2_Y -2(1-Y)^3 \partial_Y]\psi|_{Y=0}
\end{align}
assuming the spherical symmetry.  Its discretized version reads
\bea
\nabla^2 \psi|_{r=0} \rightarrow 3\left(
\frac{\psi_{-1}+\psi_{1}-2\psi_0}{\Delta Y^2}-4\frac{\psi_{1}-\psi_{-1}}{\Delta Y}
\right)=6
\frac{\psi_{1}-\psi_0}{\Delta Y^2}
\eea
where the Neumann boundary condition is implemented by setting $\psi_{-1}=\psi_1$ to get the last equality.

In consideration of time evolution, we use Crank-Nicolson method for stability of the system. With the discrete time, \eqref{radSchEq} becomes
\begin{align}
i \frac{\psi^{k+1} -\psi^{k}}{\Delta t}= \frac{1}{2}\Bigl\{
 (-\frac{1}{2}\nabla^2\psi^{k+1}+V^{k+1}\psi^{k+1}) +(-\frac{1}{2}\nabla^2\psi^k+V^k\psi^k)\Bigr\} . \label{CN}
\end{align}
(In this equation we have suppressed spatial indices for a clear presentation of our method for time evolution.) 
Note that we take an average of two values at different time-indices $k$ and $k+1$. Supposing that we know the values of the wavefunction up to the time index $k$, \eqref{CN} implicitly determines $\psi^{k+1}$ at next time. To find an explicit form we arrange this equation
\begin{align}
\left[1-\frac{\Delta t}{2i}(-\frac{1}{2}\nabla^2+V^{k+1})\right]\psi^{k+1}=\left[1+\frac{\Delta t}{2i}(-\frac{1}{2}\nabla^2+V^{k})\right]\psi^{k} . 
\label{CN1}
\end{align}
In the case of ordinary linear Schr\"odinger equation, the potential $V$ is time independent and then we might directly obtain $\psi^{k+1}$ by inverting $\left[1-\frac{\Delta t}{2i}(-\frac{1}{2}\nabla^2+V)\right]$. But in our case, the potential $V^k$ is a solution of \eqref{radLaplacian} depending on $\psi^k$ and thus $V^{k+1}$ involves unknown function $\psi^{k+1}$. So we determine $\psi^{k+1}$ in an iterative procedure. Let us approximate $V^{k+1}$ by $V^k$ at initial stage, then we may obtain an approximate value of $\psi^{k+1}$ by inverting  $\left[1-\frac{\Delta t}{2i}(-\frac{1}{2}\nabla^2+V^k)\right]$. 
With this $\psi^{k+1}$, we solve the second equation \eqref{radLapEq} to find a improved value of $V^{k+1}$.
In the second stage, we may obtain an improved value for $\psi^{k+1}$ by inverting the left hand side of \eqref{CN1} with the improved  $V^{k+1}$  obtained in the previous step.
By repeating this procedure, we can find very accurate value for $\psi^{k+1}$.

Below we shall be mainly interested in the probability density distribution in $Y$ space given by
\bea
\rho_Y(Y,t) = 4\pi  \frac{ Y^2}{(1-Y)^4} |\psi(Y,t)|^2 .
\eea
The time dependent flow of the probability density will give us a detailed information on the dynamics of Schr\"{o}dinger-Newton system. Further we shall be interested in a rotation 
curve defined by the relation
\bea
v(r,t)=\sqrt{\frac{G {\cal M}(r,t)}{r}}
\eea
where ${\cal M}(r,t)$ is defined by the integrated mass within the radius $r$  that is given by
\bea
{\cal M}(r,t) =4\pi M \int_0^r dr' {r'}^2 |\psi(r',t)|^2 .
\eea
\begin{figure}[htb]
\centering{\resizebox{3.2in}{!}{\includegraphics{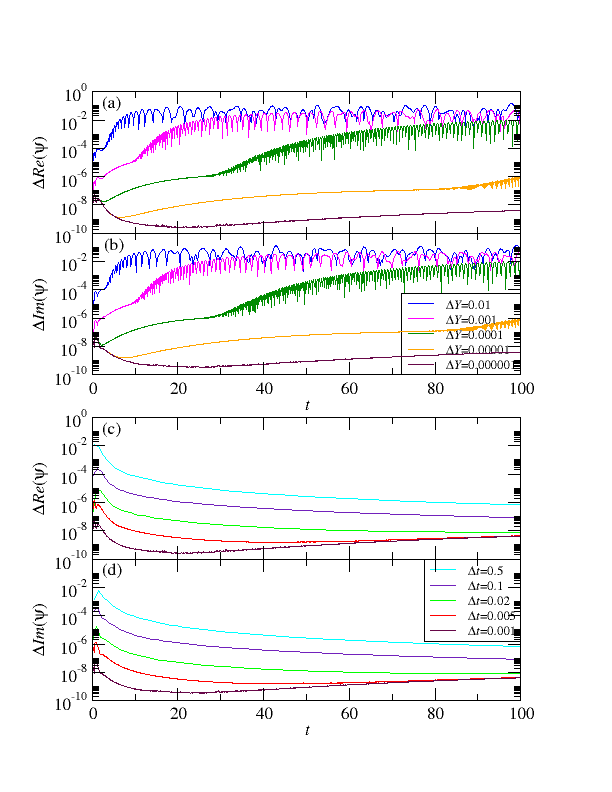}}}
\caption{
Mean absolute error in the wavefunction using the 
Crank-Nicolson method as a function of time (see text), for (a) 
$Re(\psi_{I})$, $\Delta t$=$10^{-3}$, (b) $Im(\psi_{I})$, 
$\Delta t$=$10^{-3}$, (c) $Re(\psi_{I})$, $\Delta Y$=$10^{-6}$, and 
(d) $Im(\psi_{I})$, $\Delta Y$=$10^{-6}$. The initial wavefunction we 
use is of the form $\psi_{I}(r, 0)=\pi^{-3/4}e^{-\frac{1}{2} r^2}$, 
and $F$=0.}
\label{fig-accuracy}
\end{figure}
\begin{figure}[htb]
\centering{\resizebox{3.2in}{!}{\includegraphics{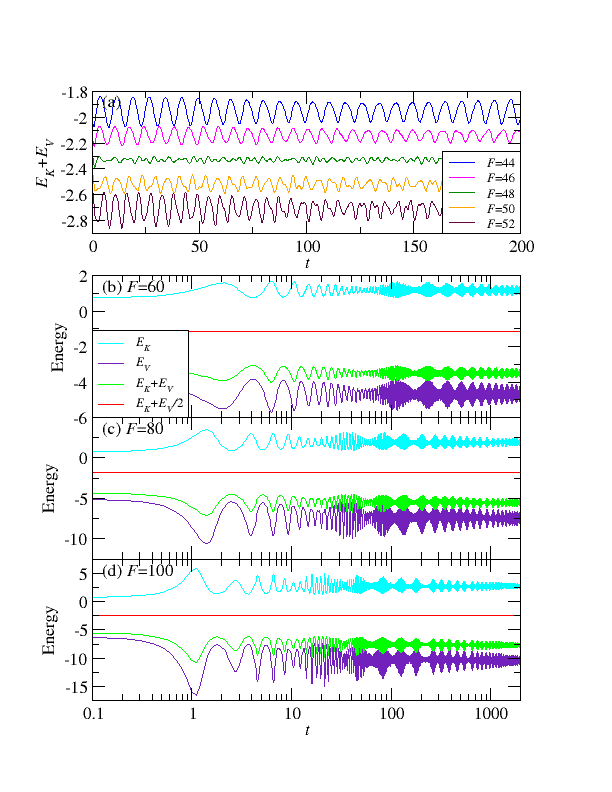}}}
\caption{ (a) $E_K$+$E_V$ for the initial wavefunction of the form 
$\psi_{I}(r,0)=e^{-r^2/2}/\pi^{3/4}$ as a function of time and the 
parameter $F$. A stationary state occurs at $F=48$. The energy as a 
function of time for
(b) $F=60$, (c) $F=80$, and (d) $F=100$. 
In all 
panels, $E_K$ denotes the kinetic energy, $E_V$ the potential energy, 
$E_K$+$E_V$ the Hamiltonian  energy, and $E_K$+$E_V$/2 the conserved energy.
Here $\Delta t$=$10^{-3}$ and $\Delta Y$=$10^{-6}$.}
\label{fig-eone}
\end{figure}
\begin{figure}[htb]
\centering{\resizebox{3.2in}{!}{\includegraphics{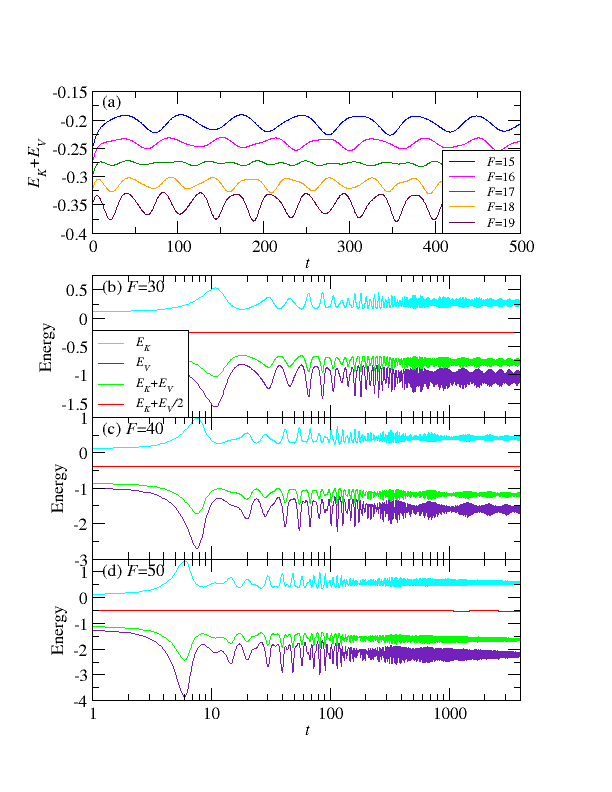}}}
\caption{(a) $E_K$+$E_V$ for the initial wavefunction of the form 
$\psi_{II}(r,0)=e^{-r/2}/\sqrt{8\pi}$ as a function of time and the 
parameter $F$. A stationary state occurs at $F=17$. The energy as a 
function of time for 
(b) $F=30$, (c) $F=40$, and (d) $F=50$. 
In all 
panels, $E_K$ denotes the kinetic energy, $E_V$ the potential energy,
$E_K$+$E_V$ the Hamiltonian energy, and $E_K$+$E_V$/2 the conserved energy.
Here $\Delta t$=$10^{-3}$ and $\Delta Y$=$10^{-6}$.}
\label{fig-etwo}
\end{figure}

\section{Results}
In this section we present various types of benchmarks of the
Crank-Nicolson 
method, which verify that the self-consistent 
optimization scheme we employ performs efficiently for a wide range
of discretization parameters. We also demonstrate that the use of the
self-consistent technique has potential advantages such as favorable
scaling in reducing the step sizes in both time and space, and the 
ease with which can be applied to more complicated nonlinear equations.

{
Figure~\ref{fig-accuracy} summarizes the performance of the 
self-consistent Crank-Nicolson 
method for a non-interacting 
Schr\"{o}dinger-Newton equation as a function of time and the step sizes 
in both space and time. At $F=0$ the model is analytically tractable 
if the initial wavefunction is of the form of the first type.
In Figures~\ref{fig-accuracy}(a) and (c) ((b) and (d)), we show how the 
mean absolute error of the real (imaginary) part in the wavefunction, 
defined as $\Delta\psi=\sum_i^{N_Y}|\psi_i^{CN}-\psi_i^{exact}|$, changes 
as a function of time and the step size in time $\Delta t$ 
(space $\Delta Y$).
As can be seen in 
Figures~\ref{fig-accuracy}(a) and (b), for large 
spatial step sizes we find an oscillatory behavior in the error 
corresponding to numerical instability, which characterizes the 
deterministic nature of the numerical algorithm as well as the exact 
treatment of the boundary condition at infinity. Importantly, the 
decrease of the step size in space would alleviate the problem
of numerical stability on finite difference discretization.
 As shown in Figures~\ref{fig-accuracy}(a) and (b),
the error of our method decreases as a power law in step sizes.

{
The complete energy results, summarized in Figures~\ref{fig-eone} and 
\ref{fig-etwo} are notable. In all cases, while the calculated Hamiltonian
energy corresponding to the expectation of the Hamiltonian $\langle H\rangle$ does not 
preserve in time, the  total conserved energy $\mathcal{E}$ to be gained from the 
gravitational interaction is independent of time as was mentioned previously.
Figure~\ref{fig-eone}(a) (\ref{fig-etwo}(a)) shows the total energy in 
the wavefunction of the first (second) type as a function of time for 
a region 
44 $\leq F \leq$ 52 (15 $\leq F \leq$ 19), where an almost stationary 
state occurs for $F$ nearly at 48 (17). The amplitude of energies 
grows with increasing the magnitude of $F$, and 
decreases as time is increased.
Our numerical results demonstrate that the period strongly depends on 
the magnitude of $F$ as well as initially excited states settle into 
stable ground state together with the asymptotic streamline of particles.}

\begin{figure}[th!]
\begin{center}
\includegraphics[width=11cm]{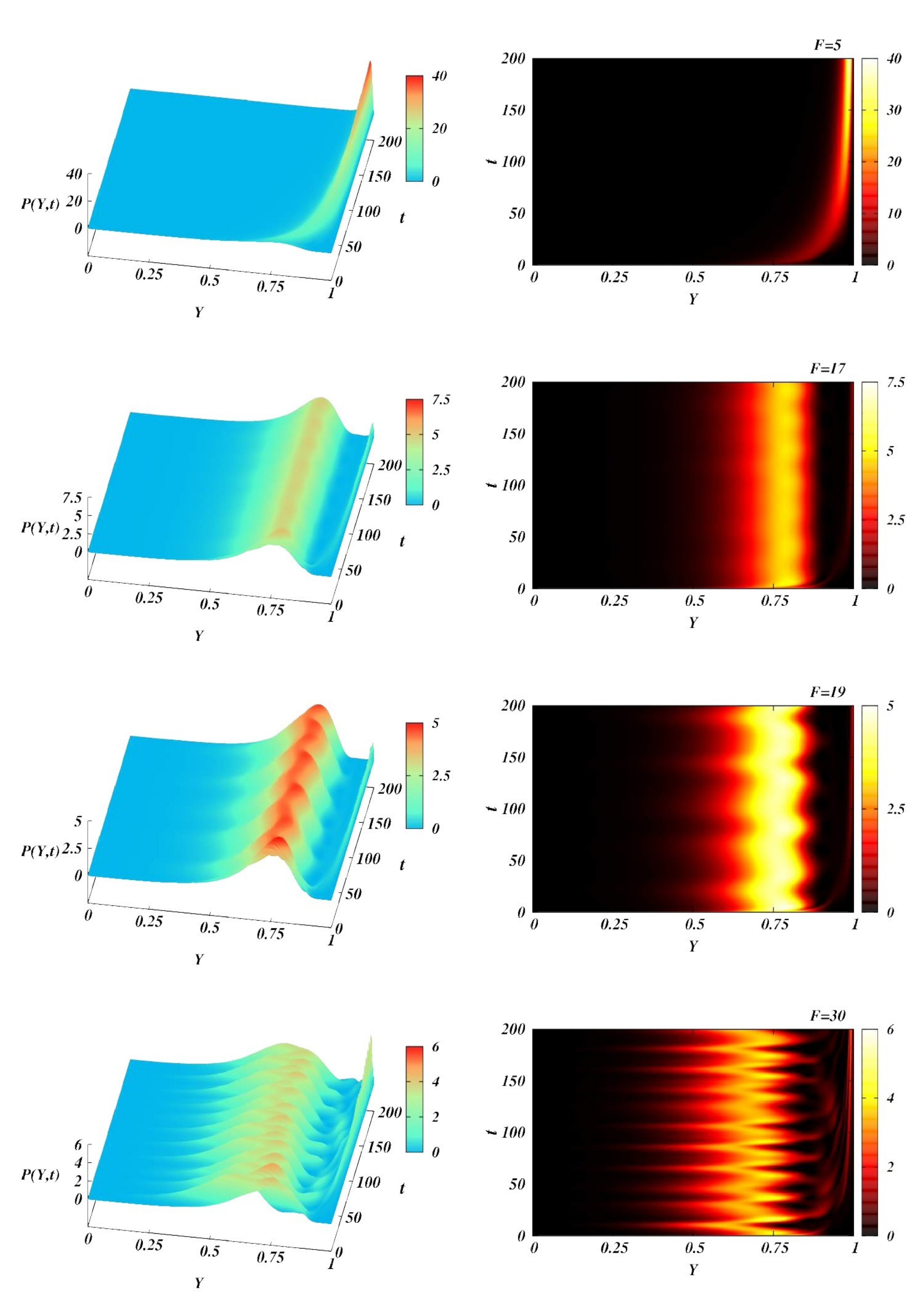}
\end{center}
\caption{
\label{figII001} The probability densities in $Y$ space are depicted for the values of 
$F=5,17,19,$ 
and $30$ 
(from the top to the bottom row) with the initial condition $\psi_{II}$. }
\end{figure}
\begin{figure}[th!]
\begin{center}
\includegraphics[width=16.2cm]{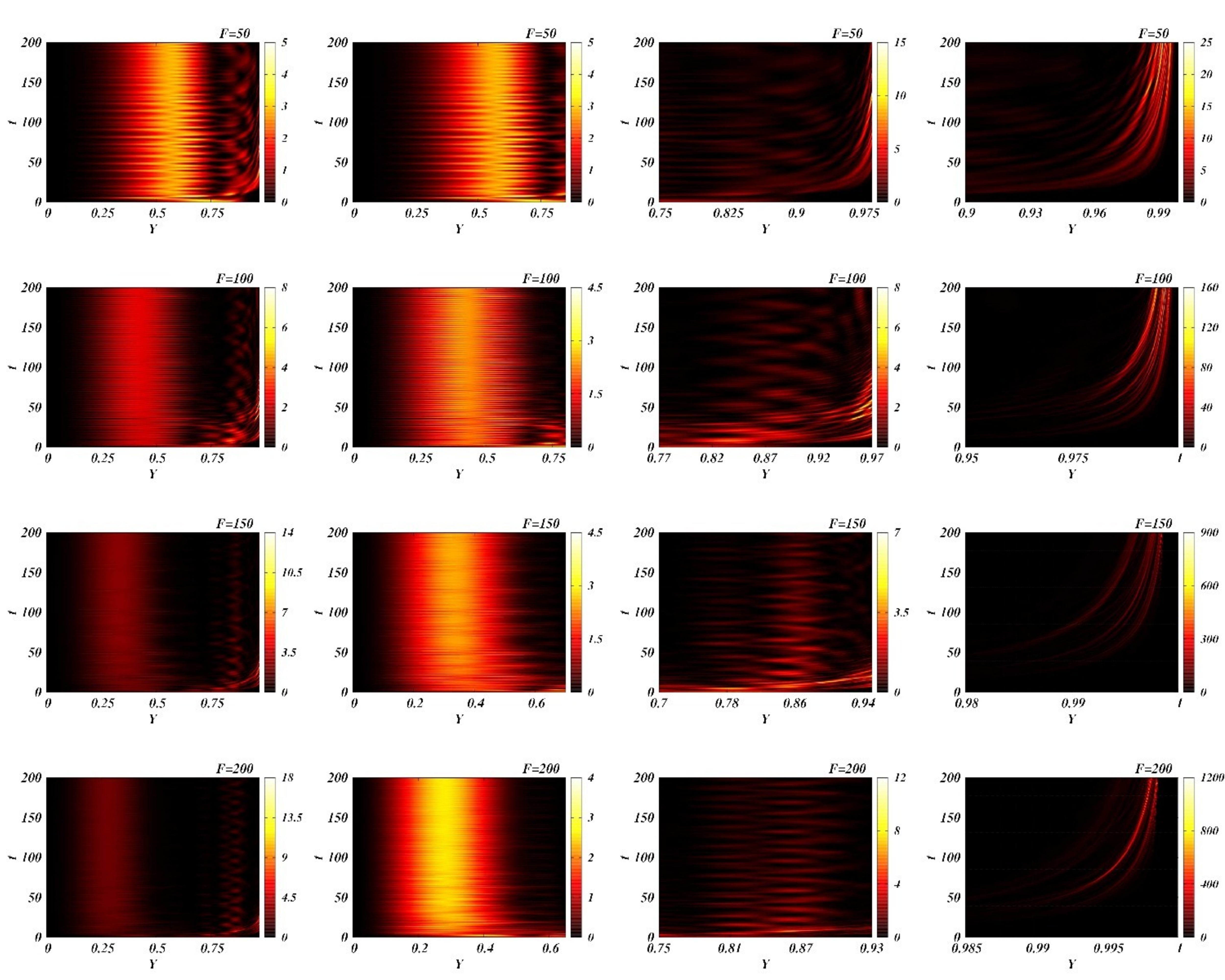}
\end{center}
\caption{
\label{figII002} The probability densities in $Y$ space are depicted for the values of 
$F=50,100,150,$ 
and $200$ from the top line to the bottom  with the 
initial condition $\psi_{II}$. 
The first column is for the full region of $Y \in [0,1)$ and the second and the third are zooming the solitonic core and excited states respectively.
One can see the escaping streaming of probability current to the asymptotic region in the fourth column.
}
\end{figure}
\begin{figure}[th!]
\begin{center}
\includegraphics[width=11cm]{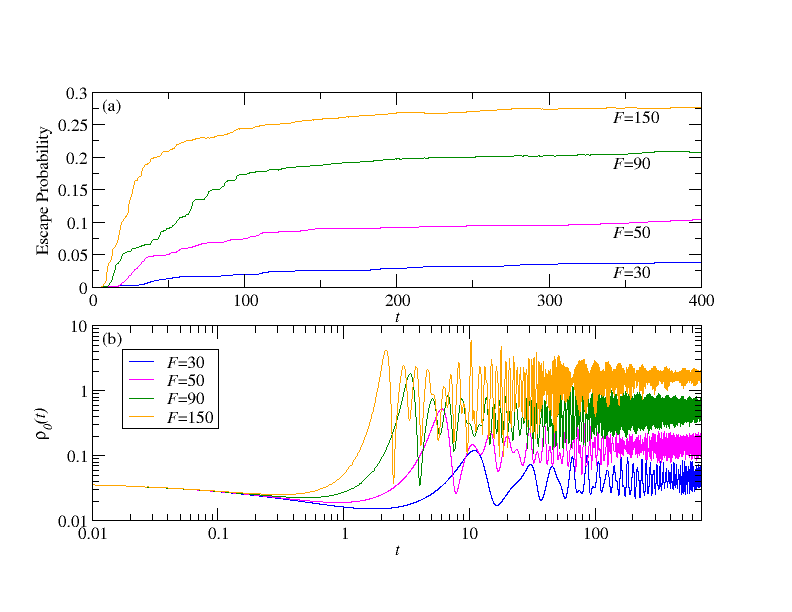}
\end{center}
\caption{
\label{figII004} We present the escaping probabilities in the region beyond $r = 24$ (or $Y =
0.96$) on the left and the values of $\rho_0(t)=|\psi(0,t)|^2$ at the origin as functions of time, for the cases
$F = 30,\, 50,\,90,\,$
and
$150$. 
}
\end{figure}
 \begin{figure}[th!]
\begin{center}
\includegraphics[width=8.8cm]{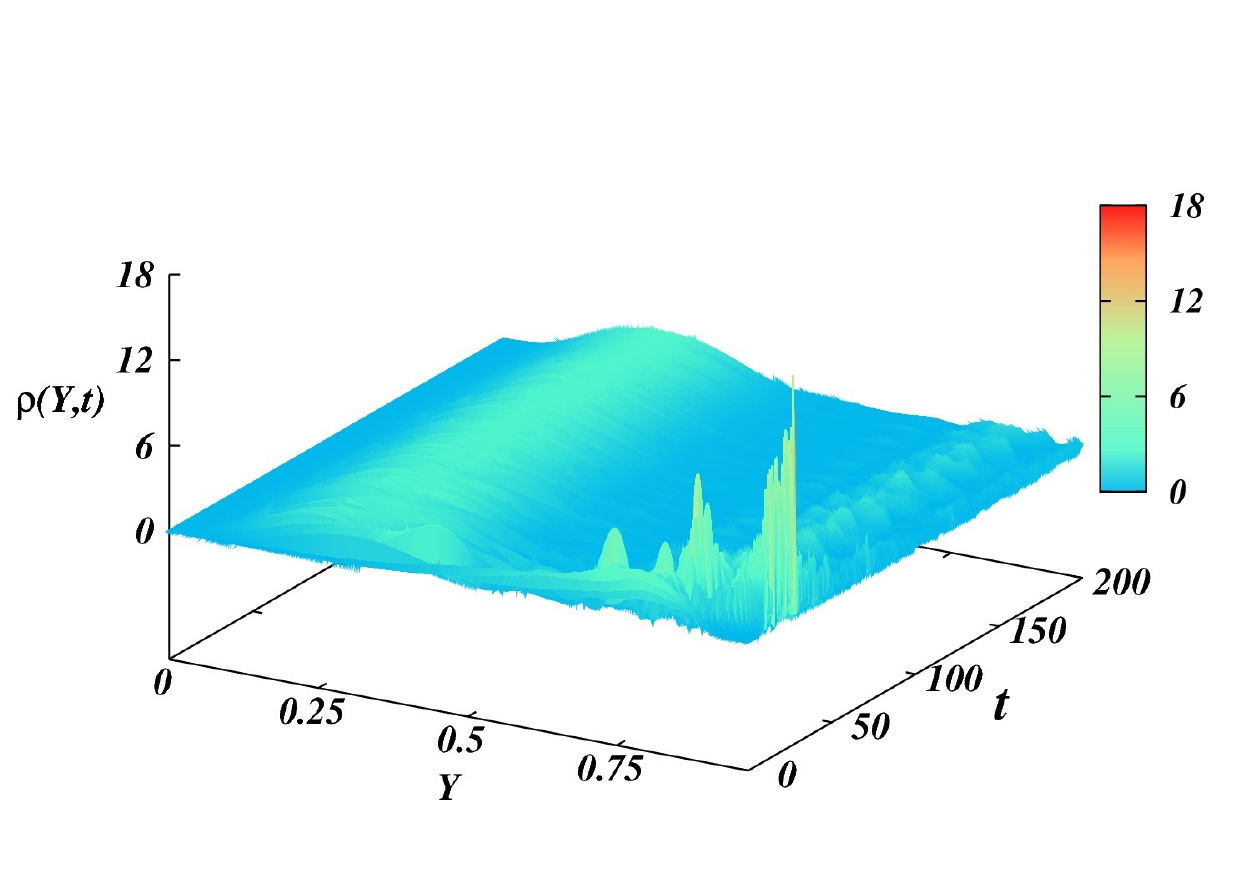}\includegraphics[width=7.6cm]{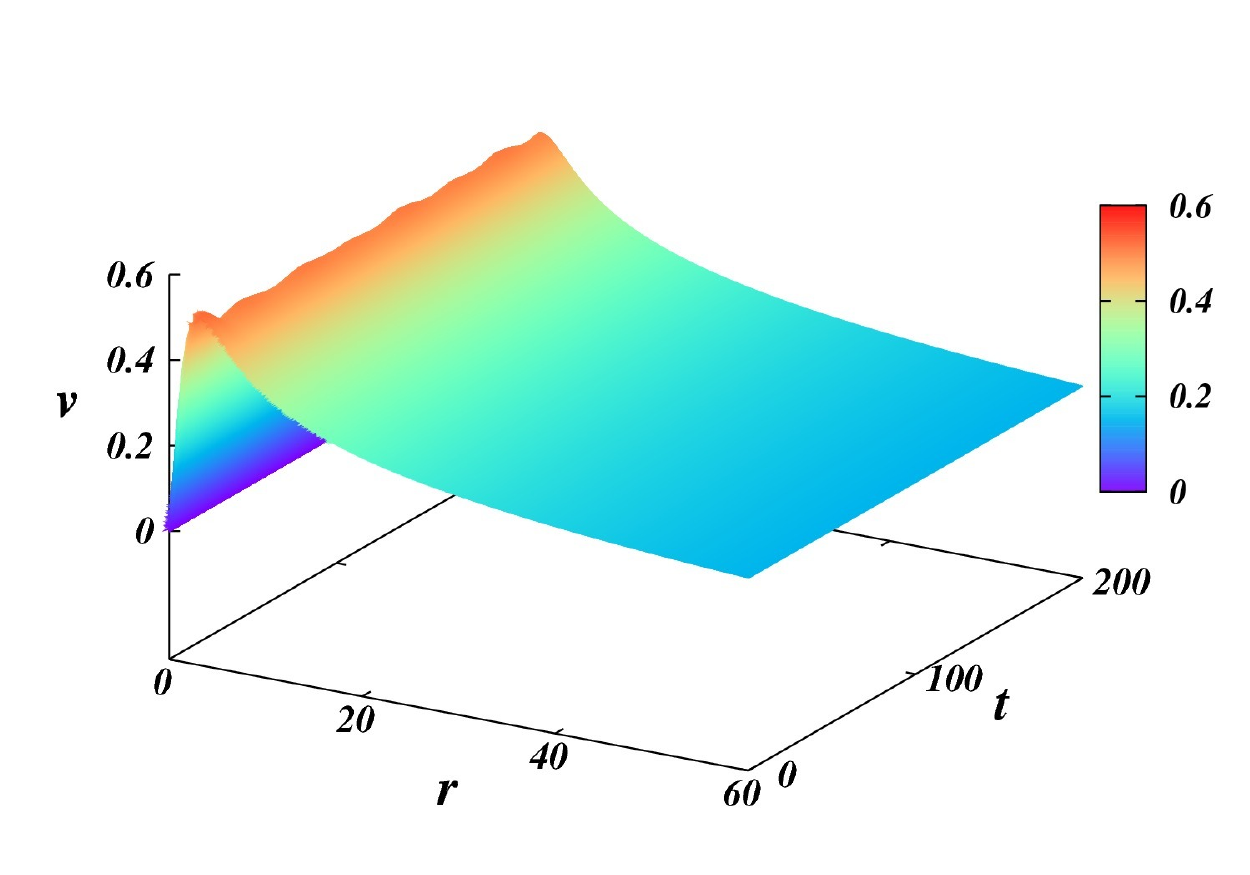}
\end{center}
\caption{
\label{figII005} On the left the probability density in $Y$ space is depicted as 3D diagram for  $F=200$ with the initial condition $\psi_{II}$. One can see 
the escaping streaming of probability current to the asymptotic region. On the right, we draw the rotation curve $v$ for $F=17$.
}
\end{figure}

 Let us  now present our main results of numerical analysis. Here in this presentation, we shall use the initial condition $\psi_{II}$ given in (\ref{initial}). Our primary focus
  is the time development of the probability density $\rho_Y (Y,t)$ in the $Y$ space. Let us first illustrate its  general trends as the parameter 
$F$ (or the total mass $M$) changes. 
  
  If $F$ is small enough, there is in general a  negligible  gravitational attraction compared to the kinetic part of energy. Ignoring the potential contribution, the system follows the dynamics of free Schr\"{o}dinger system  where eventually all the probability will spread out completely as $t$ goes to infinity. This feature is clearly seen in the first low of Figure \ref{figII001},  which is for $F=5$.  There is no  small-sized left-over core state while most of the probability spreads out to asymptotic regions.
 
 When $F$ is around $F_0 \sim 17$, there is a change of the above trend as mentioned previously. Our initial configuration $\psi_{II}$ is rather close to that of the stable solitonic core state which is the lowest energy eigenstate of the
corresponding   Schr\"{o}dinger-Newton system. This stable soliton core state is the remnant state after completion of the gravitational cooling. $F_0$ in the above is defined by the relation
$\langle r \rangle_{\psi_{II}} =\langle r \rangle_{\rm soliton-core}(F_0)$ where one equates the expectation value of $r$ with the initial state $\psi_{{II}}$ with the expectation value of $r$ of the solitonic core state as a function of  $F$. When $F=F_0$, one finds that the escape probability  is minimized and the state remains almost stationary as was shown the second row of Figure \ref{figII001}.
 
 When $F$ is larger than $F_0$,  the gravitational attraction of the initial configuration is big enough to form a small-sized solitonic core state as one can see in Figures \ref{figII001}, and \ref{figII002}. 
 Further significant fraction of particles are escaping to asymptotic regions, which is reflecting basically the phenomenon of gravitational cooling.
 In this regime, we depict a little detailed feature in Figure \ref{figII002} for $F=50,\,\,100,\,\,
150,\,\,$ and $200$ up to $t=200$. One can see that a small-sized solitonic core is forming rapidly
 whose average location in $Y$ coordinate gets smaller as $F$ becomes larger and larger. This is because the effective gravitational attractions 
become
larger as $F$ gets 
 larger. Interestingly, this core is rapidly oscillating with multiple number of frequencies reflecting nonlinear nature of our  Schr\"{o}dinger-Newton system. One sees that part of rapid 
 oscillation around peak radial position is escaping to larger distance forming the number of excited states. These excited states are unstable and decay completely escaping to 
 infinity eventually. Such feature is depicted in   Figure \ref{figII002}. 
 In the figures in the first column of  Figure \ref{figII002}, we draw 
$\rho_Y(Y,t)$ for the entire range of $Y$.  The figures in the second/third column 
illustrate
solitonic cores/excited halo states. The figures in the fourth column
are for the streamlined flow of probability escaping  to asymptotic regions.
In the left panel of Figure \ref{figII005}, we also show the 3D diagram of 
$\rho_Y(Y,t)$ for $F=200$ and  
 one can see the feature of gravitational cooling in a rather clear manner.
 
Figure \ref{figII004}(a) shows 
the escape probability $P(r > 24)$ as a function of time  which is an integrated probability of finding particles in the region beyond  $r = 24$. 
 We choose   $F=30,\,\,50,\,\,90$ and $150$.
There is always an initial time 
 delay since the initial flow of particles needs time to reach $r=24$. 
 As seen in Figure \ref{figII004}(a), the maximum time 
delay for $F=150$ is approximately 10.
After the delay, there is rather a stiff growth of probability in time which saturates to a certain 
 value eventually.  The full escape probability is around $0.38$ when $F=200$ for instance. 
As shown in Figure \ref{figII004}(a), the escape probability 
is strongly enhanced as $F$ increases.


Figure \ref{figII004}(b) also shows
the central density $\rho_0(t)=|\psi(0,t)|^2$ 
 for $F=30,\,\, 50, \,\,90, \,\,$ and $150$, 
hence, one finds they are oscillating rapidly as found in the corresponding solitonic cores
 and settle down to certain values eventually as the gravitational cooling progresses. 

Figure \ref{figII005} emphasizes the peculiar features of the system 
that have been found in our numerical calculations.
The left panel of Figure \ref{figII005} shows the escaping streaming 
of probability current to the asymptotic region in the presence of 
strong gravity for $F=200$ $(\gg F_0)$, confirming that gravity 
interactions strongly affect the characteristic of gravitational 
cooling.
In the right panel of Figure \ref{figII005}, we further show the 
rotation curve $v$ as functions of $t$ and $r$ which again displays 
the stationary nature of $F=17$ configuration as the probability in 
Figure \ref{fig-etwo}.

\section{Discussions}
In this note, we numerically study time evolution of the spherically 
 symmetric Schr\"{o}dinger-Newton system
using the self-consistent Crank-Nicolson approach. Compactification of radial 
 coordinate enables us to investigate gravitational cooling patterns in a 
 realiable way. It is shown that the numerical errors are successfully 
 controlled. 
 Our simulation clearly shows that a small-sized solitionic core forms 
 accompanied with number of excited halo states and a large fraction of 
 particles escape to asymptotic regions. Besides these common features, it 
 also shows how the patterns of these gravitational coolings depend on the 
 values of total mass. 

 Even if our work is limited to spherically symmetric cases in this work, 
 it should be interesting to study possible angular structures of 
 Schr\"{o}dinger-Newton system with or without axial symmetry. This will be 
 our next work. 
 Another interesting direction of research is numerical simulation of very 
 large structures in super-galactic scale. Recent simulations assumed cold 
 dark matter. It is very intriguing to see the results of simulations 
 involving self gravitating Schr\"{o}dinger field instead of cold dark 
 matter  \cite{Schive:2014dra} in a rather detailed manner.
 
\section*{Acknowledgement}
D.B. was
supported in part  by 2018 Research Fund of University of Seoul.

\end{document}